\documentclass[twocolumn,nofootinbib,amsmath,amssymb,a4paper]{revtex4}

\usepackage{graphicx}
\usepackage{dcolumn}
\usepackage{bm}

\def\bea{\begin{eqnarray}}
\def\eea{\end{eqnarray}}
\def\beq{\begin{equation}}
\def\eeq{\end{equation}}
\def\cbeta{ \cos{2 \beta}}
\def\sbeta{\sin{2 \beta}}

\def\bdbarp{\hbox{$B_d$\kern-1.4em\raise1.4ex\hbox{\barp}}}
\def\bbarp{\hbox{$B$\kern-1.1em\raise1.4ex\hbox{\barp}}}
\def\sss{\scriptscriptstyle}
\def\ks{K_{\sss S}}

\def\Act{{\cal A}_{ct}}
\def\Aut{{\cal A}_{ut}}

\def\cbeta{ \cos{2 \beta}}
\def\sbeta{\sin{2 \beta}}

\def\dt{\Delta t}
\def\kspipi{\ks\pi^+\pi^-}

\def\dm{\Delta m}

\begin{document}

\title{ $\beta(\phi_1)$ from $ B $ Decays to Charm}

\author{Alakabha Datta}
 \email{datta@phy.olemiss.edu}
\affiliation{%
Department of Physics and Astronomy \\
University of Mississippi\\
Oxford, MS 38677, USA\\
}%

\begin{abstract} 
The measurement of $\sin {2\beta}$ from $B_d \to J/\psi K_s$ does not resolve the discrete ambiguities in
the angle $\beta$. A measurement of $\cos {2\beta}$ is therefore desirable.
This talk is about measuring the CKM angle 
$\beta( \phi_1)$ from various 
 $B$ decays to charm final states which also allow us a measurement of
 $\cos {2\beta}$. 
\end{abstract}

\maketitle

\section{Introduction} We already have a precise measurement of
$\sin {2\beta}$ from $B_d \to J/\psi K_s$. However this measurement does not resolve the 4 fold ambiguity in $\beta$:
$$ ( \beta, { \pi \over 2} - \beta) \quad
( \beta + \pi, { \pi \over 2} - \beta + \pi) .$$
We want to measure the angle $\beta$  in many 
processes to test the SM. Also measurement of both $\sin {2\beta}$ and
$\cos {2\beta}$ is clearly desirable to partly resolve the discrete ambiguity in $\beta$.

In  this talk I will review the theory of measuring the CKM angle 
$\beta( \phi_1)$ from various 
 $B$ decays to charm final states.
 I will concentrate on $B \to J/\psi K_s^{(*)} $, $B \to D^{(*)}\bar{D}^{(*)} K_s$. These are $ b \to c \bar{c} s$ transitions.
I will then discuss
 $B \to D^{(*)}\bar{D}^{(*)}$( $b \to c \bar{c} d$) and
$B \to D^{(*)}h^0$( $b \to c \bar{u} d$). There are other approaches to measuring
$\cos{2\beta}$ that will not be discussed in this paper \cite{quim}.

\section{$B \to J/\psi K_s^{(*)} $}
The decay $B \to J/\psi K_s^{(*)} $ is a VV decay and there are three amplitudes-
  $A_{0}$, $A_{\|}$ which are CP even and $A_{\perp}$ which is CP odd.
 The time dependent angular distribution allows us to measure both $\sbeta$ and $\cbeta$. The time independent angular distributions is needed to fix the coefficients of  in the time dependent distributions \cite{prd1}.
 However the time independent analysis gives the triple products, \cite{dattaTP}
\begin{eqnarray}
Im (A_\parallel ^*A_\perp) &=& |A_\parallel ||A_\perp| \sin(\delta_\perp - \delta_\parallel ), \nonumber\\
Im (A_\perp A_0^*)         &=& |A_\perp|     |A_0|     \sin(\delta_\perp-\delta_0),
\end{eqnarray}
but the coefficient of $\cbeta$ term depend on the cosine of the phase differences $\delta_{\perp} - \delta_{\parallel}$ and
$\delta_{\perp} - \delta_{0}$. To resolve this ambiguity in the coefficient term information is obtained from the decay
 $B(t) \to J/\psi K \pi$.
The $K \pi$ system originating from $B\to J/\psi K \pi$ can, in principle, have any integer spin but it is found experimentally that,
below 1.3  GeV, the $S$ and $P$ waves dominate.
Assuming that the strong interactions between the $J/\psi$ and the $ K \pi$
are small, one can obtain additional information about the P-wave phase shift by an angular analysis of the decay $B\to J/\psi K \pi$. This additional information can then be used to resolve the sign ambiguity in $\cbeta$ measurement.
The result of such an analysis yields a positive value for $\cbeta$ \cite{prd1}, which combined with that the value of $\sbeta$ obtained in
 $B_d \to J/\psi K_s$ results in a value of $\beta$ which is consistent with Standard Model expectations.

\section{ $B(t) \to D^{*+}D^{*-} K_s$}
This decay can have both non resonant and resonant contributions.
The resonant contributions can go through an intermediate excited $D^{**}$
making this process sensitive to $\cbeta$ measurement \cite{browder}.

 We define the following amplitudes
\bea
{a^{\lambda_1,\lambda_2}}  & \equiv &
 A(B^0(p)\to D^{+*}_{\lambda_1}(p_{+}) D^{-*}_{\lambda_2}(p_{-}) K_s(p_k)),\
 \eea
 \bea
{\bar a^{\lambda_1,\lambda_2}} & \equiv & 
A(\bar B^0(p)\to D^{+*}_{\lambda_1}(p_{+}) D^{-*}_{\lambda_2}(p_{-}) K_s(p_k  ),\
\label{AAbar}
\eea
where $B^0$ and $\bar  B^0$ represent unmixed neutral $B$ and $\lambda_1$ and $\lambda_2$ are the polarization indices of the $D^{*+}$ and $D^{*-}$ respectively.
\begin{figure}
\includegraphics[width=0.4\textwidth]{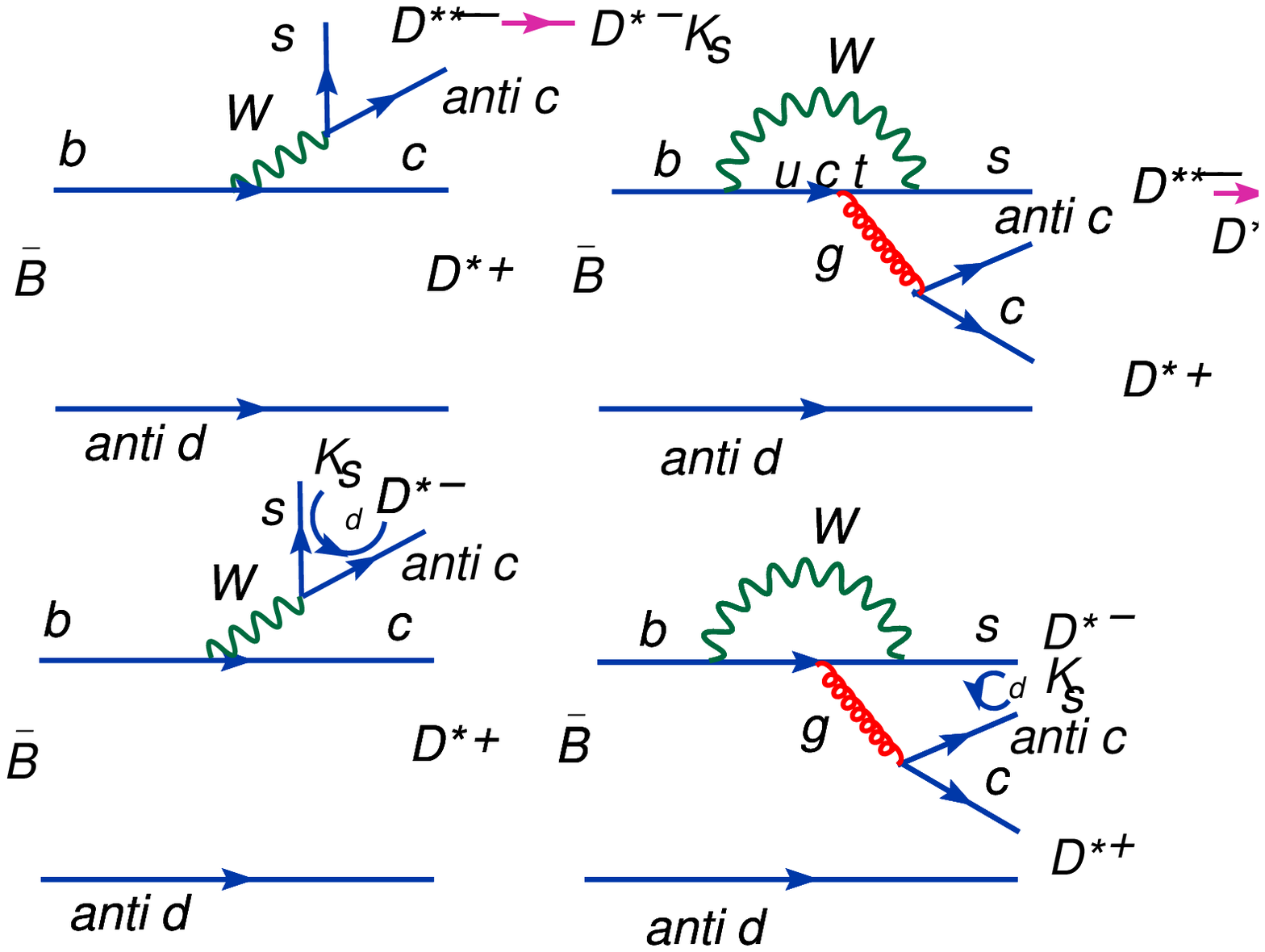}
\caption{\label{fig:narrow1} The decay $ B \to D^{*+}D^{*-} K_s $.}
\end{figure}
The time-dependent amplitudes for an oscillating state $B^0(t)$ which has been tagged as a $B^0$ meson at time $t=0$ is given by 
\bea
A^{\lambda_1,\lambda_2}(t)& =& {a}^{\lambda_1,\lambda_2} \cos\left(\frac{\Delta m\,t}2\right) + i e^{-2i \beta} \nonumber\\
& &{\bar{a}}^{\lambda_1,\lambda_2} \sin\left(\frac{\Delta m\,t}2\right),\
\eea
and the time-dependent amplitude squared summed over polarizations and integrated over the phase space angles is:
\bea
|A(s^+,s^-;t)|^2 & = & \frac{1}{2}\left[ {\rm G}_0(s^+,s^-)+{\rm G}_{\rm c}(s^+,s^-)\cos\Delta m\,t- \right.\nonumber\\
& & \left.{\rm G}_{\rm s}(s^+,s^-)\sin\Delta m\,t \right], \nonumber\\ 
{\rm G}_0(s^+,s^-)     & = & |{a}(s^+,s^-)|^2 +|\bar{{a}}(s^+,s^-)|^2 ,\\
{\rm G}_{\rm c}(s^+,s^-) & = & |{a}(s^+,s^-)|^2 -|\bar{{a}}(s^+,s^-)|^2 ,\\
{\rm G}_{\rm s}(s^+,s^-) & = & 2\Im\left (e^{-2i \beta} \bar{a}(s^+,s^-)
{{a}^\ast(s^+,s^-)} \right ) \nonumber \\
& = & -2 \sin(2\beta)\, \Re \left ( \bar{a}{{a}^\ast} \right ) + 2\cos (2\beta) \,\Im \left ( \bar{a}{{a}^\ast} \right ),\nonumber\\
 s^+ & = & (p_{+} +p_k)^2,\quad s^-= (p_{-} +p_k)^2,
\eea
where
$ s^+= (p_{+} +p_k)^2,\quad s^-= (p_{-} +p_k)^2$. It is convenient to
replace 
$ (s^+, s^-)  \rightarrow (y, E_k) $
where $E_k$ is the $K_s$ energy in the rest 
frame of the $B$. The variable $y$ is defined as,
$y=\cos{\theta}$ with $\theta$ being the angle between 
the momentum of $K_s$ and $D^{*+}$ in a frame where the two $ D^{*}$ are moving back to back along the z- axis. Note that, $s^+\leftrightarrow s^-$ corresponds to $y \leftrightarrow -y$.
\begin{figure}
\includegraphics[width=0.4\textwidth]{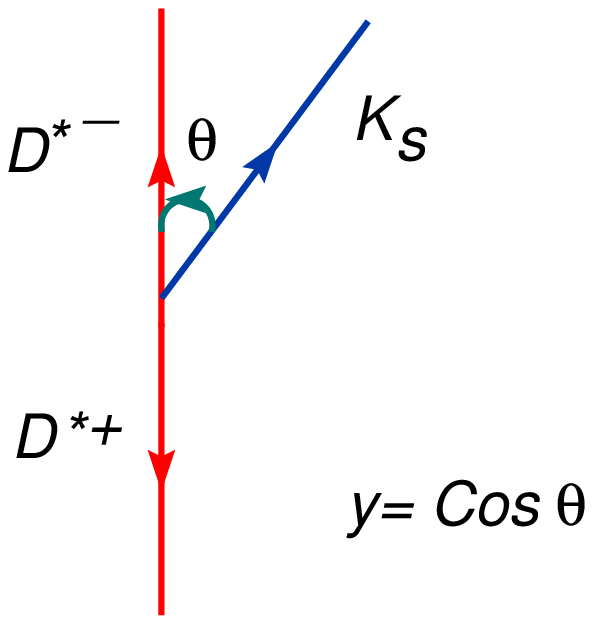}
\caption{\label{fig:narrow2} The definition of the variable $y$.}
\end{figure}
Carrying out the integration over the phase space variables $y$ and $E_k$ one 
gets the following expressions for the time-dependent total rates
for $ B^0(t) \to D^{*+}D^{*-}K_s$ and the CP conjugate process,
\bea
\Gamma(t) &= &\frac{1}{2}[I_0 + 2 \sin( 2\beta)\sin (\Delta m t) I_{s1}],\\
{\overline{\Gamma}}(t) &= &\frac{1}{2}[I_0 - 2 \sin( 2\beta)
\sin (\Delta m t) I_{s1}],\
\eea
where $I_0$ and $I_{s1}$ are the integrated $G_0(y,E_k)$ and 
$G_{s1}(y,E_k)$ functions.
One can then extract $\sin (2 \beta)$ from the rate asymmetry
\beq
\frac{\Gamma(t)-{\overline{\Gamma}}(t)}{\Gamma(t)+{\overline{\Gamma}}(t)} =
D\sin( 2\beta)\sin (\Delta m t);\quad
D = \frac{2I_{s1}}{I_0}, \
\eeq
where $D$  is the dilution factor. Furthermore,
\beq
I_0= 2\Gamma(0)=2{\overline{\Gamma}}(0).\
\label{I0}
\eeq

 The $\cos(2 \beta)$ term can be probed by integrating over half the 
range of the variable $y$ which can be taken for instance to be $y \ge 0$ for
$B$ decay and $y \le 0$ for ${\bar {B}}$ decay. 
\bea
\hspace{-15.0mm}\frac{{\Gamma}(t, y \ge 0)}{I_0} = \quad \quad \quad\ \quad \quad & &  \nonumber\\
 \frac{1}{2}[\frac{{J}_0}{I_0} 
+\frac{{J}_c}{I_0} \cos (\Delta m t)
 + 2 \sin( 2\beta)\sin (\Delta m t) 
\frac{{J}_{s1}}{I_0}\nonumber\\
 -
2 \cos( 2\beta)\sin (\Delta m t) \frac{{J}_{s2}}{I_0}],\\
\frac{{\overline{\Gamma}}(t, y \le 0)}{I_0} = 
\quad \quad \quad\ \quad \quad & &  \nonumber\\   \frac{1}{2}[\frac{{J}_0}{I_0} 
+\frac{{J}_c}{I_0} \cos (\Delta m t) - 2 \sin( 2\beta)\sin (\Delta m t) 
\frac{{J}_{s1}}{I_0}\nonumber\\
 -
2 \cos( 2\beta)\sin (\Delta m t) \frac{{J}_{s2}}{I_0}],\
\eea
where ${J}_0$, ${J}_c$, ${J}_{s1}$ and  ${J}_{s2}$,  
are the integrated $G_0(y,E_k)$, $G_c(y,E_k)$, $G_{s1}(y,E_k)$ and 
$G_{s2}(y,E_k)$ functions integrated over the range $y \ge 0$.
One can measure $\cos(2 \beta)$  by fitting to the time distribution
of ${\Gamma(t) +{ \bar \Gamma}(t) \over I_0}$.

The fits to the data yield \cite{babarDDK}
\bea
\frac{J_c}{J_0} &=& 0.76\pm 0.18(stat)\pm 0.07(syst), \nonumber \\
\frac{2J_{s1}}{J_0}\sbeta &=& 0.10\pm 0.24 (stat) \pm 0.06 (syst), \nonumber \\
\frac{2J_{s2}}{J_0}\cbeta &=& 0.38\pm 0.24 (stat) \pm 0.05 (syst). \
\eea
Using the theoretical calculation of $\frac{2J_{s2}}{J_0}$ \cite{browder}
$\cbeta$ is preferred to be positive at the 94\,\% confidence level
as the theoretical parameter $J_{s2}/J_0$ is positive.  It is also interesting to note that
$J_c/J_0$ is significantly different from zero. This implies
that there is a sizable resonant contribution to the 
decay  from a unknown $D^+_{s1}$ state with large
width. This can have important implications for the interpretation of the
new $D_{sJ}$ states discovered recently \cite{dattapat}.

\section{ $B(t) \to D h^0 $ }
The decay $B(t) \to D h^0 $ has been recently proposed to measure
$\sbeta$ and $\cbeta$ \cite{bondar}. The idea in this case is also to use intermediate resonances to probe $\cbeta$. Consider the decays of the $B$ meson to $D h^0(h^0=\pi^0, \eta..)$.
The dominant decay is through the $b\to c\bar{u}d$ quark level process.
\begin{figure}
\includegraphics[width=0.4\textwidth]{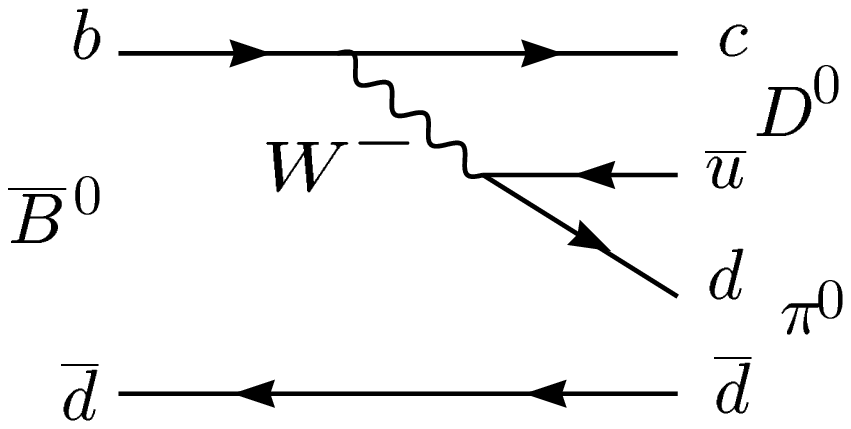}
\caption{\label{fig:narrow3} The decay $B \to D h^0 $ .}
\end{figure} 
Next consider the multi body $D$ meson decay.
As an example use $D \to K_s \pi^+ \pi^-$. We can write,
\bea 
Amp(\bar{D}^0 \to \kspipi)  & = & f(m_+^2,m_-^2),  \nonumber\\
 Amp({D}^0 \to \kspipi)  & = & f(m_{-}^2,m_{+}^2),  \nonumber\\
{m_{+,-}}^2=(p_{K_s}+p_{\pi^{\pm}})^2.\
\label{ampD0}
\eea
 The amplitude for the $B$ decay at time $t_{\rm sig}$ is then given by,
\bea
  M_{\bar{B}^0}(\dt) & = &
  [\bar{B}^0 \to D^0(K_s \pi^+ \pi^-) h^0] C  - \nonumber\\ 
 & & i \frac{p}{q} \eta_{h^0} (-1)^l [{B}^0 \to \bar{D}^0(K_s \pi^+ \pi^-) h^0] S, \nonumber\\
 \ 
\eea
where $C=\cos(\dm\dt/2)$ and $S=\sin(\dm\dt/2)$ and
\bea
  \label{eq:m_b0bar_pre}
  M_{\bar{B}^0}(\dt) & = &
  f(m_-^2,m_+^2) C - \nonumber\\ 
 & & i \frac{p}{q} \eta_{h^0} (-1)^l f(m_+^2,m_-^2) S.
\eea
Then, using Eq.~\ref{ampD0},
\bea
  \label{eq:m_b0bar}
  M_{\bar{B}^0}(\dt) & = &
  f(m_-^2,m_+^2) C - \nonumber\\ 
 & & i e^{-i2\phi_1} \eta_{h^0} (-1)^l f(m_+^2,m_-^2) S, \\
  \label{eq:m_b0}
  M_{B^0}(\dt) & = &
  f(m_+^2,m_-^2) C - \nonumber\\ 
 & & i e^{+i2\phi_1} \eta_{h^0} (-1)^l f(m_-^2,m_+^2) S.
\eea
One then needs to model $f(m_+^2,m_-^2)$ to extract
the phase $2\phi_1(2\beta)$. This is done by assuming Breit-Wigner form for the various contributing resonances. A non resonant component is also included. The experimental measurements of this process is discussed in
 other talks in the workshop\cite{kathy}.
  
\section{$\bar{B}(t) \to D^{(*)}\bar{D}^{(*)}$}
\begin{figure}
\includegraphics[width=0.4\textwidth]{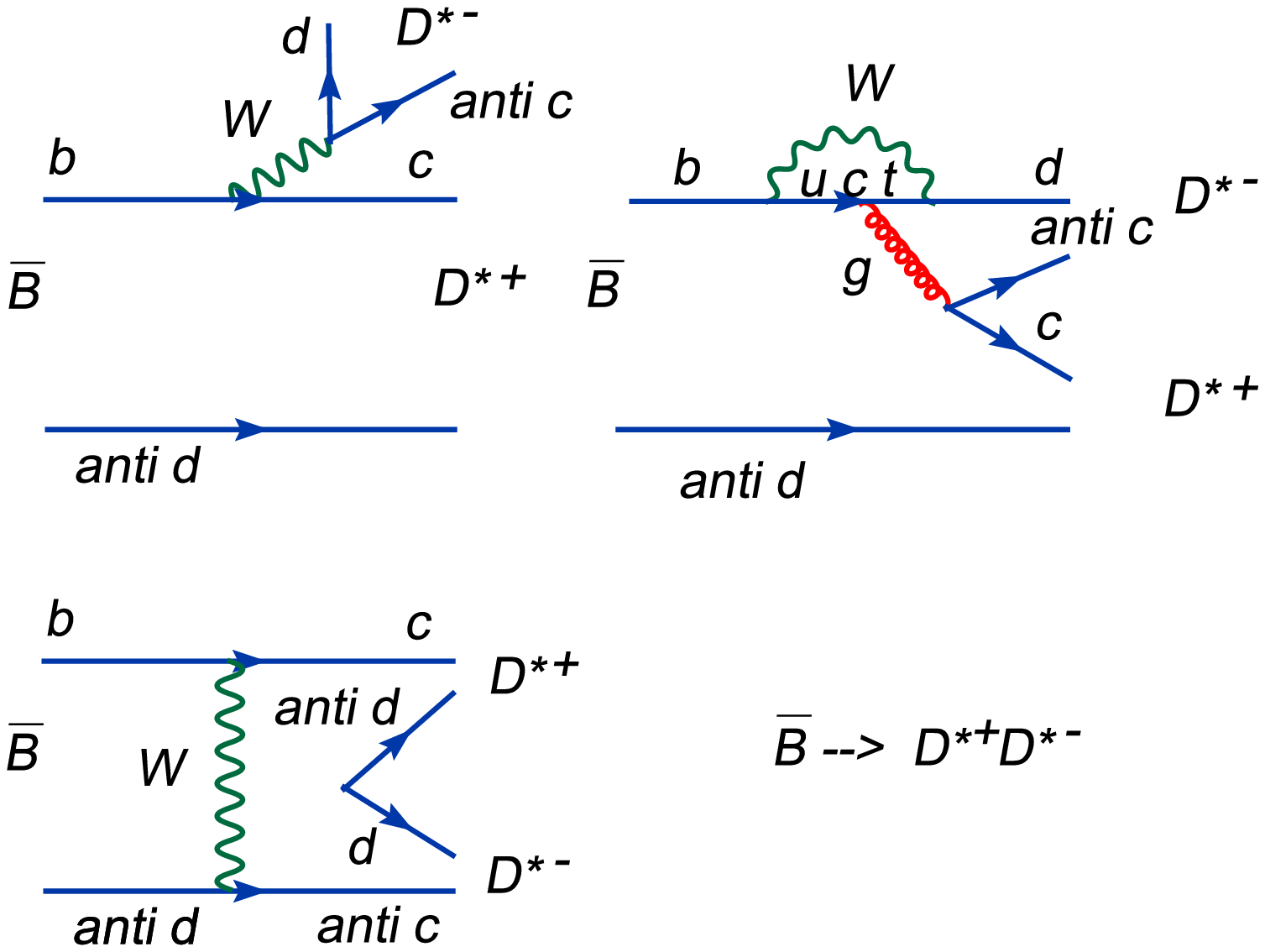}
\caption{\label{fig:narrow4} The decay $\bar{B} \to D^{(*)}\bar{D}^{(*)}$ .}
\end{figure} 
The amplitude for $\bar{B} \to D^{(*)}\bar{D}^{(*)} $ can be written as,
\bea
A^D & = & (T + E + P_c)  V_{cb}^* V_{cd} + P_u  V_{ub}^* V_{ud} + \nonumber\\
& &(P_t + P_{EW}^C)  V_{tb}^* V_{td}, \nonumber\\
& = & (T + E + P_c - P_t - P_{EW}^C)  V_{cb}^* V_{cd} \nonumber\\
& & + (P_u - P_t -
P_{EW}^C)  V_{ub}^* V_{ud}, \nonumber\\
& \equiv & \Act\ e^{i \delta^{ct}} + \Aut\ e^{i \gamma} e^{i
\delta^{ut}}.
\label{AmpDDbar}
\eea 
The time-dependent measurement of this decay allows one to
obtain the three observables,
\bea
B &\equiv & \frac{1}{2} \left( |A|^2 + |{\overline A}|^2 \right) = 
\Act^2 + \Aut^2 \nonumber\\
& + & 2 \Act \, \Aut \cos\delta \cos\gamma , \nonumber\\
a_{dir} &\equiv & \frac{1}{2} \left( |A|^2 - |{\overline A}|^2 \right)
= - 2 \Act \, \Aut \sin\delta \sin\gamma ,\nonumber\\
a_{indir} &\equiv & {\rm Im}\left( e^{-2i \beta} A^* {\overline A} \right) \nonumber\\
&- &\Act^2 \sin 2\beta 
 -  2 \Act \, \Aut \cos\delta \sin (2 \beta + \gamma)
\nonumber\\
&-& \Aut^2 \sin (2\beta + 2 \gamma). \
\eea
%
The three independent observables depend on five theoretical
parameters: $\Aut$, $\Act$, $\delta$, $\beta$, $\gamma$. If $\Aut$ can be neglected then one can obtain $\sbeta$. In general, however, 
 one
cannot obtain CP phase information from these measurements- this is the well known penguin pollution problem. Hence theoretical input is necessary to get the CKM phase 
information. One can partially solve for the theory amplitudes as, 
\bea
\Act^2 & = & { a_R \cos(2\beta + 2\gamma) - a_{indir} \sin(2\beta +
2\gamma) - B \over \cos 2\gamma - 1}, \nonumber\\ 
a_R^2 & = & B^2 - a_{dir}^2 - a_{indir}^2 ~.
\eea
One can then   obtain $\Act$ from a partner process and use it as a theory input. We can then obtain $\gamma$ if given $\beta$ and vice versa.
The partner process can be chosen to be 
 $ \bar{B} \to {D}^{(*)}\bar{D}_s^{(*)}$ \cite{justin} or 
 $ \bar{B_s} \to {D}_s^{(*)}\bar{D}_s^{(*)}$ \cite{fleischer}.

\section{Conclusions}
The measurement of both $\sbeta$ and $\cbeta$ is crucial to test the standard model predictions of CP violation. This is all the more important
as there are several decays where there are hints of beyond standard model physics\cite{dattanp}, making the  precise measurements
of the CKM angles a top priority. In this talk we have described several decays involving B decays to charm that probe both $\sbeta$ and $\cbeta$.

\begin{acknowledgments}
 I thank the High Energy Group of the University of Mississippi to partially fund my trip to the workshop.
\end{acknowledgments}

\end{document}